# Affinity and Hostility in Divided Communities: a Mathematical Model

Chris Thron and Rachel McCoy

Department of Sciences and Mathematics, Texas A&M University-Central Texas
1001 Leadership Place, Killeen TX USA 76549

thron@tamuct.edu

## *Abstract*

We propose, develop, and analyze a mathematical model of intergroup attitudes in a community that is divided between two distinct social groups (which may be distinguished by religion, ethnicity, or some other socially distinguishing factor). The model is based on very simple premises that are both intuitive and justified by sociological research. We investigate the behavior of the model in various special cases, for various model configurations. We discuss the stability of the model, and the continuous or discontinuous dependence of model behavior on various parameters. Finally, we discuss possible implications for strategies to improve intergroup affinity, and to defuse tension and prevent deterioration of intergroup relationships.

## 1       Purpose and Scope

Many local communities, especially in the Third World, are markedly divided between two or more very distinct ethnic or religious groups. In sub-Saharan Africa for instance, many villages, towns, and cities comprise distinct groups of "Christian" and "Muslim" residents. In some cases, residential neighborhoods are completely integrated; in others, districts or quarters belong to one or the other religion. These situations may or may not be accompanied by interreligious tensions (Rasmussen, 1993; Mehler, 2005, 127).  In other situations, towns may be divided between two dominant ethnicities, such as Hutu and Tutsi in Burundi and Rwanda (Dominus, 2014). In Kenya, many local and regional conflicts have pitted two resident tribes against each other (Kimenyi, 2005, 126).

Divided communities are not unique to Africa.  Following the Bosnian war, cities in Bosnia had physical boundaries that divided different ethnicities (Bosnian Serbs, Bosniacs, and Bosnian Croats), ensuring that the different ethnic groups did not mix in daily life or work.  Over time, some grassroots efforts have been initiated to increase contacts between the two groups (Demichelis, 1998, 3). "Peace walls" have separated Catholic and Protestant neighborhoods in Northern Ireland for more than forty years: residents generally acknowledge that these walls have helped to preserve safety and security, but have also negatively impacted quality of life (Byrne et al., 2012).

On many occasions, horrific violence has resulted when relations between co-resident groups has degenerated. Even when outright violence is avoided, hostility, mistrust, and avoidance between neighboring factions have considerable economic and social costs, which serve as strong motivations for improved relations. A better understanding of the dynamics of affinity/hostility between groups may be invaluable in fostering amicable relationships.

This paper proposes and analyzes a heuristic mathematical model of intergroup relations within a community consisting of two distinct groups clearly identified by religion or ethnicity. Mathematical modeling is an increasingly important tool in shaping social policy. Although mathematical models have significant limitations and are prone to oversimplification, nonetheless they can provide valuable assistance in understanding, predicting, preventing, and resolving adverse social conditions.  We shall use our model to address the following questions:

- What type of evolution can be expected for different model conditions (which are exemplified by different model parameter values)?

- What stable distributions of affinity/hostility are possible for different model conditions?

- Do stable distributions vary continuously with changes in conditions, or are there cases where small changes in conditions produce large differences in the distribution of affinities/animosities?

- What strategies that target model conditions are sufficient to bring about desired changes in intergroup attitudes?
- What model conditions may produce deterioration in intergroup attitudes?

Our model is based on a small number of intuitive principles, and contains only a few parameters. The values of these parameters may not be measurable in a practical situation: but even without knowing the parameters' values, the model can be used to predict qualitative behavior of a split community. This information may be potentially useful in devising strategies to reconcile groups and promote stability.

The remainder of the paper is structured as follows. In Section 2 we survey the current literature on mathematical models of intergroup violence. In Section 3 we discuss examples of programs designed to mitigate tensions between divided communities. In Section 4 we present the heuristic justification of the model, and its mathematical specification. In Section 5 we provide mathematical analyses of the model under various particular conditions. In Section 6 we present several characteristic simulations and their implications for model behavior. In Section 7 we summarize our findings and draw conclusions. Section 8 lists references.

# 2 Survey of mathematical models of intergroup conflict

## 2.1 Statistical studies

Several authors have produced empirical statistical analyses of violence on a national scale. Some representative examples are as follows:

- Elbsdawi and Sambanis (2000) consider the question of why there are so many civil wars in Africa. They perform an empirical statistical analysis on the likelihood of civil war in African nations, as well as nations in other continents. It is suggested that Africa is more likely to have civil wars not because of ethnic and religious variations, but because of the lack of solid government and economics. Factors such as poverty level, economics predominantly based on natural resources, and low education levels are cited as contributing factors to likelihood of civil wars. They claim that to create a stable environment, Africa needs to use its diversity (through cooperative programs) as an asset, and needs democratic political reform paired with more economic programs that are not solely based on natural resources.

- Collier and Hoeffler (2002) use an empirical/analytical model to show that Africa's civil wars can be predicted based on economic performance. They hypothesize that all governments have citizens who are unsatisfied and willing to resort to violence. The probability that this will happen depends on the rebels' ability to financially raise an army of equal size to the government's military force. The more financially stable a country is, the more sizable a country's defense, and the less likely a rebel force can financially support a military rebellion. Also related to economic stability, Collier and Hoeffler found that the lower per capita income and higher population rate increases the risk of conflict.

## 2.2 Agent-based models

Alternatively, some researchers use agent-based models in an attempt to understand civil violence. Some representative examples are given as follows:

- Schelling (1969) presents an agent-based model which does not address violence per se, but rather the evolution of communities that consist of two coexisting groups with separate ethnic identities. The model's structure is very simple: The community is modeled as a checkerboard, and the residences belonging to the different ethnicities (which are the "agents" in the model) are represented by markers of two different colors. The markers move according to rules designed to reflect individuals' relative preference to live next to others like themselves. Schelling demonstrates that under these conditions, segregated communities are self-forming: integrated communities will decompose into segregated regions. The severity of segregation depends on the parameters set for each group of agents. The higher the demand of a group to be around others like themselves, the more segregated and tightly clustered the neighborhood will become. If one group has a higher tolerance for neighbors unlike themselves, then the community will have the more tolerant group spread widely with small clusters of the more intolerant group packed tightly into areas of the community. These eventual results are relatively independent of initial conditions, but do depend on system parameters. Although Schelling's model initially received little attention, it gained considerable prominence in the U.S. during the debate over residential segregation in the 1980s and 1990s (Clark and Fossett, 2006).

- Epstein (2002) created a model of the spread of social violence which similarly uses agents on a checkerboard. The model employs population agents of three types—actively violent, inactive, or jailed—as well as "cop" agents that are able to "arrest" nearby active agents. The likelihood of an agent becoming actively violent depends on that agents' sense of grievance, as well as the risk of being arrested (which depends on the presence of "cops" in the vicinity). Epstein shows that under certain conditions localized bursts of violence can occur. He suggests that the model may give insight into the dynamics behind outbursts of interethnic violence, and that this insight could provide valuable insight into the development of policies to prevent and contain violence.

- Salerno et al. (2011, 2013) developed the National Operational Environment Model (NOEM), which includes an agent-based model of population behavior on a national scale. NOEM is intended to supply decision makers with information about possible social responses (including civil violence) to military operations that affect infrastructure and other physical conditions. The population response model is fundamentally based on Epstein's model.

- Thron et al. (2012) show that an agent-based model can duplicate the statistical characteristics exhibited by service protests in the Gauteng Providence in South Africa from 2004-2010. Starting with a simplified version of the NOEM as basis, adjustments are made to include media influence, which accounts for the fact that violence does not always spread in a geographically continuous manner.

- Thron and Jackson (2013) consider the general application of Epstein-style agent-based models of social violence to practical situations. They argue that the model does not accurately reflect spatial mechanisms that lead to the formation of social violence, and that the model is simply displaying generic features common to systems that exhibit self-organized criticality. They conclude that the model is better suited as a phenomenological description rather than a dynamical explanation of social violence capable of prediction.

- Nizamani et.al (2014) models the relationship between public outrage and public violence. Using five types of agents (sensitive, upset, violent, relaxed, and immune), a model is constructed based on a system of differential equations. Nizamani applies this model to analyze the effects that an event has on populations in other parts of the world (for example a movie made in one country that causes public outrage in another). In this model, all agents start out as "sensitive" with only a few having knowledge of the event. The agents with knowledge will be randomly designated as either "immune" or "upset". Public outrage then spreads by a mathematical mechanism which models word-of-mouth propagation. The model produces a pattern of outrage that can turn to violence, but always ends with either all "immune" or "relaxed" agents. The severity of the outrage depends on the parameters used for the agents.

## 2.3 Other mathematical models

There is also literature on mathematical models of coalitions and strategies in situations of inter-group violence:

- Zachary (1977) develops an information flow model, which he applies to the case study of a karate club that experience conflict which eventually resulted in the club's breaking apart. Using information about members' opinions and interactions with each other outside of the club meetings, Zachary was able to explain most members' choice of affiliation following the breakup.

- The text of Jones (2000) describes game theory models and strategies for conflict. She models cooperative games, non-cooperative games, zero sum games, matrix games, and bargaining games.

- Azam (2001) investigates the role of government in redistribution of resources as a strategy for mitigating conflict. Azam builds a theoretical model that analyzes the chance of rebellion due to different government actions. His model shows that appropriate redistribution of resources and aid may discourage rebellion; while inequitable distribution of aid can actually incite rebellion. An important conclusion is that ethnic identity should not be stamped out, but used to help build up a country.

- Pawlak (2004) develops a mathematical representation of intergroup conflict, using the current conflict in the Middle East as an example. Countries in the Middle East are modeled are represented as nodes on a graph. For each issue, each edge of the graph is assigned a value between -1 and 1, where value of -1(1) reflects complete disagreement (agreement) between the two agent-nodes joined by the given edge. Based on this information, Pawlak defines functions that characterize degree of conflict, which Pawlak argues can be used as an aid to decision-making and in the mathematical modeling of conflicts.

- Fraenkel and Grofman (2004) use a mathematical model to evaluate the effectiveness of Horowitz's alternative voting scheme (Horowitz, 2003) in producing moderate politics in electorates that are deeply divided along ethnic or religious lines. The alternative voting scheme attempts to give an advantage to parties that can attract votes across the divide. They supplement their analysis with an examination of the 1999 elections in Fiji, and conclude that practical results vary greatly from Horowitz's prediction.

- Osborne (2010) hypothesizes that conflict between ethnic groups occurs when it is economically more profitable. He considers an asymmetrical case of two ethnic groups where one group (the majority) has all the power to control taxation of resource production, while the other group (the minority) can choose to either to share resources and submit to taxation or move towards separatism and share resources internally. The minority group's choice is not all or nothing, but rather partial sharing of resources is also allowed by the model. The model predicts that the amount of resources shared by the minority varies inversely with the amount of taxation. Osborne also performs an empirical analysis of intergroup conflict, taking into account factors such as distribution of the population, ordinal ranking according to the World Bank's ease of doing business scale, and percent of the country's GDP that consists of food. He gives tables showing which variables were significant at 10%, 5%, and 1% levels. These empirical findings confirm his hypothesis that ethnic conflict should increase when incentive for economic cooperation decreases. He cites economic freedom as an important component of peaceful interethnic relations.

# 3    Strategies for mitigating tensions in divided communities:

One strategy that has been used to reduce tensions in divided communities is to encourage positive contact between groups through sports. Some examples are as follows:

- The NGO Mercy Corps used this strategy in Sri Lanka to address the hostility and mistrust between Tamils, Sinhalese, and Muslims following an extended, ethnically-based civil war. In 2011 over a period of 12 months, Mercy Corps conducted workshops for youth leaders and coaches, and organized multi-ethnic sporting events. The program involved two thousand youth from twelve villages, and Mercy Corps reported a 20% increase in positive relations across groups ("Peace and reconciliation," 2011).

- The Football 4 Peace program, initiated by the University of Brighton's School of Sport and Service Management, trains coaches to instill cooperation and provides opportunities for children from different groups to play on the same team. The original program was in Israel, involving Jewish and Arab children; subsequent programs have been conducted in Jordan and Ireland/Northern Ireland ("Football 4 Peace", n.d.).

- Right to Play International, headquartered in Canada, is active in about 20 countries around the world and aims at teaching cooperation and acceptance to children in violence-prone communities through games and providing an alternative outlet to guns and violence ("Right to Play", n.d.).

Similar examples include the Open Fun Football Schools (involving 22 countries in Eastern Europe and central Asia) ("Open Fun", n.d.), and a number of sports-based initiatives in post-apartheid South Africa. An analysis of these South African initiatives by Höglund and Sundberg leads the authors to conclude that poorly-organized programs can actually enhance conflict between groups, but well-run initiatives can serve both as a bridge between divided communities and as an outlet for entertainment to keep violent behavior from developing (Höglund and Sundberg, 2008). An analysis of several reconciliation-through-sports initiatives by Selliaas, leads him to the following conclusions (Selliaas, 2007):

- It is hard to draw firm conclusions concerning the effectiveness of these programs, as they are relatively new;

- All of these programs are started after some measure of stability has been achieved;
- NGO's play an important role in the implementation of these programs;
- Sports do not start reconciliation, but can play a role in a bigger reconciliation project.

# 4    Mathematical model specifications

Our mathematical model is based on common-sense observations about human interactions that are also supported by sociological research. We first state these principles as heuristic "rules of thumb", and then provide corresponding mathematical translations.

## *4.1    Heuristic formulation of model*

The following "rules of thumb" are assumed to govern interpersonal interactions in a community consisting of two strongly distinct social groups.

H1) Individuals within each group have varying degrees of affinity towards the other group, which can be ranked on a linear scale. On the one end are the extremists, who despise and avoid the other group and are prone to inciting violence. On the other end are the moderates, who treat individuals from the other group on an almost equal basis.

H2) Constructive (cooperative) interactions between individuals in different groups tend to improve their affinities for each other, due to the mutual benefit derived from the interaction. Such contacts may occur through daily commerce, education, community development, sports, and so on. This assumption is empirically supported by the outcomes of the sports programs described in Section 3.

H3) Isolation and lack of contact between groups tend to increase hostility: separatism leads to rumor-mongering, mistrust, and misunderstanding (Bourhis, 2009, 46). On the other hand, more contact leads to more accommodation. For example, in the Zanzibar region of Tanzania, ninety eight percent of the population is Muslim: but as tourism and Western influence has increased, strict rules for behavior during Ramadan become more relaxed and behaviors are not strictly enforced (Lodhi, 1994, 92).

H4) Extremists in either group will tend to avoid and discourage interactions with the other group. (Hardin, 2002)

H5) In interpersonal interactions, individuals tend to influence other individuals towards their own opinion. (Vallacher & Nowak, 2005)

## *4.2    Mathematical formulation:*

The heuristic assumptions (H1)–(H5) have been translated into mathematical specifications as follows:

M1) The community consists of two groups of equal size, each group modeled as a set of *N* agents.

M2) Each individual in each group has an *affinity* towards the other group. Any individual's affinity is a number between 0 (extreme hostility) and 1 (the other group is treated on equal basis with the individual's own group). At the beginning, each individual's affinity is chosen randomly between 0 and 1. We will denote the affinity of agent *A* as $a(A)$. This specification reflects the heuristic assumption (H1).

M3) Over the course of the simulation, individuals interact with each other, and their affinities change as a result of the interaction. For simplicity, we assume that individual interactions involve just two agents, and that interactions occur sequentially one after the other. Each interaction is modeled as follows:

   a. Choose a random agent $A_1$ who participates in the interaction.
   b. Choose a second agent $A_2$ as follows. With probability $1 - a(A_1)/2$, the second agent is in the same group as $A_1$. Otherwise, the second agent is chosen randomly from the other group. (This rule reflects assumption (H4), since it implies that the frequency of interaction with the other group is positively correlated with affinity.) Note that $a(A_1)=1$ implies the second agent is chosen from either group with equal probability, while $a(A_1)=0$ means that the second agent is always chosen from the individual's same group.
   c. Change the two agents' affinities based on the interaction, as follows.

If the two agents are in the same group, then:

$$a(A_1) \rightarrow a(A_1) - b_1 + c \cdot h(a(A_2) - a(A_1)) + \sigma \nu_1;$$
$$a(A_2) \rightarrow a(A_2) - b_1 - c \cdot h(a(A_2) - a(A_1)) + \sigma \nu_2.$$

If the two agents are in different groups, then:

$$a(A_1) \rightarrow a(A_1) + b_2 + c \cdot h(a(A_2) - a(A_1)) + \sigma \nu_1;$$
$$a(A_2) \rightarrow a(A_2) + b_2 - c \cdot h(a(A_2) - a(A_1)) + \sigma \nu_2.$$

In these equations,

$b_1$ denotes the (negative) drift in affinity due to a single same-group interaction.

$b_2$ denotes the (positive) drift in affinity due to a single inter-group interaction.

$c$ denotes the strength of "cohesion", that is the tendency of two interacting agents to influence each other towards their own opinion.

$h$ is a function (which we shall refer to as the "influence function") that expresses the effect of the affinity difference between two individuals on the affinity adjustment resulting from interaction between those individuals. We will consider two possibilities for the influence function: step function ($h(x) = \text{sign}(x)$) and linear ($h(x) = x$).

$\sigma$ is a noise variance.

$\nu_1, \nu_2$ are independent, identically distributed normal random variables with mean 0 and variance 1.

Since affinities are limited to lie in the range [0,1]; the values of $a$ generated by the above two equations are clipped to lie in this range: $a \rightarrow \max(0, \min(1,a))$.

The affinity change equations in M3) part (c) reflect heuristic assumptions (H2), (H3) and (H5): The $b_1$ (resp. $b_2$) terms reflect assumption (H2) (resp. (H3)) that within-group (resp. between-group) interactions tend to degrade (resp. improve) affinity towards the other group. The $c$ terms reflect (H5)'s assertion that interactions between individuals tend to bring those individuals' opinions closer together.

# 5    Theoretical Analysis of Model Behavior

## 5.1    Overview

We may gain considerable insight into the model behavior by considering various special cases. These special cases also have practical significance, which we will elucidate in each case.

In all of these special cases, we are interested in changes in composition over time of each group. In particular, we consider how the percentages of extremists and moderates changes over time; and we investigate how conditions (represented by model parameters) may be changed to tilt the balance towards increased or reduced affinity.

The special cases that we will consider are as follows:

Case 1. All individuals in each group begin with the same affinity (although the two groups' affinities may differ). In this case, we are particularly interested in the eventual steady-state affinities of the groups, depending on the initial affinities.

Case 2. Both groups are initially divided into extremist (affinity = 0) and moderate (affinity=1) factions. In this case, we are interested in the stability of configurations consisting of given proportions of extremist and moderate factions.

Case 3. Both groups start out with uniform affinity distributions on [0,1]. As in the previous case, we are interested in the eventual steady-state affinities.

In this section, we will provide approximate mathematical analyses for Case 1 and Case 2. These theoretical analyses are supplemented by the simulations presented in Section 6. In addition to these special cases, we shall derive a differential equation for the evolution of the affinity probability densities using a continuous approximation. It turns out this approximate equation can be solved exactly when the expected values of the distributions are constant. This equation gives insight into the dispersive properties of the probability distributions as a function of model parameters (especially the parameter $c$).

## 5.2  Case 1: Unanimous starting affinities for each group.

In order to gain some theoretical insight into the situation where starting affinities are unanimous for each group, we make some simplifying assumptions that enable us to compute an exact solution. Simulations performed in Section 6 will confirm that these assumptions are indeed approximately satisfied.

### 5.2.1  "Sticky model" approximation and equation

We denote by "sticky model" the case where each group's agents all have the same affinity for all time. This is never strictly accurate, but in some cases it is approximately true, and it does have the advantage that it enables an exact solution. Furthermore, we shall find that it give accurate indications of the evolution of distribution medians, even when within-group affinities do in fact diverge

The "sticky model" case may be formally specified by the following condition:

*Sticky condition*: If all agents in a particular group start out with the same affinity, then they will maintain nearly the same affinity. In other words, once the distribution of affinities for a particular group has coalesced to a single value, all the affinities move together.

The sticky condition will never be exactly true for our model: nonetheless, we may perform an idealized mathematical analysis assuming that the sticky condition is true. In this case, then the affinity of each individual in either group is equal to the mean affinity of the group to which that individual belongs.

Let $x_1$ and $x_2$ be the mean affinities for the two different groups. Then in the case where $\sigma=0$, we may calculate the changes in $x_1$ and $x_2$ due to a single interaction as follows:

I.   The probability that the agent $A_1$ is in Group 1 is ½. If this happens, there are two cases:
   a. The probability that the second agent $A_2$ is in Group 1 is $(1-x_1/2)$. In this case, $x_1$ will decrease due to the interaction by an amount $2b_1/N$, where $N$ is the total number of agents in Group 1 (which is equal to the number of agents in Group 2).
   b. The probability that the second agent $A_2$ is in Group 2 is $x_1/2$. In this case, $x_1$ will change by an amount $(b_2+c \cdot h(x_2 - x_1))/N$. Also, $x_2$ will change by an amount $(b_2-c \cdot h(x_2 - x_1))/N$.

II.  The probability that the agent $A_1$ is in Group 2 is ½. If this happens, there are two cases:
   a. The probability that the second agent $A_2$ is in Group 2 is $(1-x_2/2)$. In this case, $x_2$ will decrease due to the interaction by an amount $2b_1/N$.
   b. The probability that the second agent $A_2$ is in Group 1 is $x_2/2$. In this case, $x_2$ will change by an amount $(b_2-c \cdot h(x_2 - x_1))/N$. Also, $x_1$ will change by an amount $(b_2+c \cdot h(x_2 - x_1))/N$.

We may summarize this information in the following equations for the mean affinities after a single interaction (denoted by $x_1'$ and $x_2'$), for the case where $\sigma=0$:

$$x_1'=x_1+(1/2N)\cdot\{(1 - x_1/2)\cdot(-2b_1)+ x_1/2\cdot(b_2+ c\cdot h(x_2-x_1))\}+(1/2N)\cdot(x_2/2)\cdot(b_2 + c\cdot h(x_2-x_1)),$$

$$x_2'=x_2+(1/2N)\cdot\{(1 - x_2/2)\cdot(-2b_1)+x_2/2\cdot(b_2 - c\cdot h(x_2-x_1))\}+(1/2N)\cdot(x_1/2)\cdot(b_2 - c\cdot h(x_2-x_1)).$$

When $\sigma \neq 0$, a random noise term with variance $\sigma$ must be added to each of these equations.

In the following, we will solve this model exactly for two different functional forms of the influence function $h$. The two solutions point out different interesting mathematical aspects of the model, and also show the robustness of model behavior to changes in the functional form of $h$.

#### 5.2.1.1  Piecewise-linear version of sticky model: exact solution

If we take $h(x_2-x_1) = \text{sign}(x_2-x_1)$, then the above equations for $x_1'$ and $x_2'$ may be simplified to:

$$x_1'=x_1- (b_1/N) + (b_1/N)\cdot[ \{ 1/2 + 1/4\cdot(b_2+ s\cdot c)/b_1\}x_1+ \{ (1/4)\cdot(b_2 + s\cdot c) /b_1 \}x_2 ],$$

$$x_2' = x_2 - (b_1/N) + (b_1/N) \cdot [\ (1/4) \cdot \{\ (b_2 - s \cdot c)/b_1\} \ x_1 + \{1/2 + 1/4 \cdot (b_2 - s \cdot c)/b_1\} x_2\ ],$$

where $s \equiv \text{sign}(x_2 - x_1)$. These equations may be rewritten in matrix-vector form as

$$\mathbf{x}' = \mathbf{x} + (0.25 b_1/N)\{\ (2I + b_2/b_1 F + s \cdot c/b_1 G\ )\mathbf{x} - \mathbf{f}\ \},$$

where (using Matlab/Octave/Scilab notation for matrices, in which matrix rows are separated by semicolons)

$$\mathbf{x} = [\ x_1\ ;\ x_2];\quad F = [1,\ 1;\ 1,\ 1];\quad G = [\ 1,\ 1;\ -1,\ -1];\quad \mathbf{f} = [4;\ 4].$$

For simplicity we define

$$\beta \equiv b_2/b_1;\quad \gamma \equiv c/b_1;\quad \delta \equiv 0.25 b_1/N;\qquad A \equiv 2I + \beta \cdot F + s\gamma \cdot G.$$

Then the equation for $\mathbf{x}'$ can be written

$$\mathbf{x}' = \mathbf{x} + \delta(A\mathbf{x} - \mathbf{f}),$$

which is a linear equation. (Note however that different linear equations apply in the two regimes $x_2 > x_1$ and $x_2 < x_1$, because the value of $s$ differs.) If we let $\mathbf{k}$ denote the solution of

$$A\mathbf{k} = \mathbf{f},$$

then defining $\mathbf{z} \equiv \mathbf{x} - \mathbf{k}$ we have

$$\mathbf{z}' = (I + \delta \cdot A)\mathbf{z},$$

The solution for $\mathbf{k}$ is:

$$\mathbf{k} = 2 \cdot (1+\beta)^{-1}[1 - s\gamma\ ;\ 1 + s\gamma].$$

The vector $\mathbf{k}$ will take two values (which we denote as $\mathbf{k}_+, \mathbf{k}_-$) in the two regions $x_1 < x_2$ and $x_1 > x_2$. The two values $\mathbf{k}_+, \mathbf{k}_-$ are symmetrical with respect to the line $x_1 = x_2$: they are also symmetrical with respect to the point $(1+\beta)^{-1}[\ 2\ ;\ 2\ ]$. In the case where $b_2/b_1 = \beta = 3$, this is the point $[1/2;\ 1/2]$.

Within each region, this gives the solution for $\mathbf{x}^{(m)}$ (the value of $\mathbf{x}$ after $m$ interactions) as

$$\mathbf{x}^{(m)} = (I + \delta \cdot A)^m (\mathbf{x}_0 - \mathbf{k}_s) + \mathbf{k}_s,$$

where $\mathbf{x}_0 = (x_{10}; x_{20})$ denotes the vector of initial affinities of the system.

The eigenvectors of $A$ are the eigenvectors of $\beta F + s\gamma \cdot G$, which are $[1; -1]$ and $[1 + s\gamma; 1 - s\gamma]$. The corresponding eigenvalues are 2 and $2(1 + \beta)$. When $\delta \equiv 0.25 b_1/N$ is small (which we would expect for large populations $N$), we have the approximation:

$$\mathbf{x}^{(m)} \approx e^{tA}(\mathbf{x}_0 - \mathbf{k}_s) + \mathbf{k}_s,\quad \text{where}\quad t \equiv \delta \cdot m.$$

It follows that in this case the matrix $A$ and the vectors $\mathbf{k}_s$ ($s = \pm 1$) uniquely determine the shape of the system trajectories, including the ultimate fate of the system. The factor $\delta$ only influences how fast the system moves along each trajectory, but not the trajectory's shape. Note that $A$ and $\mathbf{k}_s$ depend only on $\beta$ and $\gamma$: so the two parameter $\beta$ and $\gamma$ effectively determine the fate of the system.

In order to generate these trajectories, we may write:

$$\mathbf{x}_0 - \mathbf{k}_s = q_1 \mathbf{v}_1 + q_2 \mathbf{v}_2,$$

where $\mathbf{v}_1$ and $\mathbf{v}_2$ are the eigenvectors $[1; -1]$ and $[1 + s\gamma; 1 - s\gamma]$, and $q_1$ and $q_2$ are constants to be solved for. Recalling that the eigenvalues are 2 and $2(1 + \beta)$ respectively, we have

$$\mathbf{x}^{(m)} \approx e^{2t} q_1 \mathbf{v}_1 + e^{2(1+\beta)t} q_2 \mathbf{v}_2 + \mathbf{k}_s.$$

From this formula, several properties of the trajectories may be derived. Since both eigenvalues are positive, it follows that the fixed points $\mathbf{k}_+, \mathbf{k}_-$ are both repellors in their respective regions. This means that trajectories will consistently move away from the location of $\mathbf{k}_s$. If $\gamma \geq \min(1, (\beta-1)/2)$, then $\mathbf{k}_s$ is located outside of the square $[0,1]^2$, and all trajectories will move towards the line $x_1 = x_2$ which separates the two regions. On the other hand, if $\gamma \leq \min(1, (\beta-1)/2)$ then only some trajectories will move towards the line $x_1 = x_2$, and others will move away. The trajectories all eventually tend to align with the direction of the second eigenvector $\mathbf{v}_2 = [1 + s\gamma;\ 1 - s\gamma]$, since the second eigenvalue has the largest magnitude.

*5.2.1.2 Nonlinear version of the sticky model*

We have noted that when the influence function $h(x_2-x_1) = \mathrm{sign}(x_2-x_1)$ then the sticky model is linear in each region $\{x_1 < x_2\}$, $\{x_1 > x_2\}$. A nonlinear version of the model is obtained when we use instead the influence function: $h(x_2-x_1) = x_2-x_1$. In this case, the evolution equations simplify to (using the previous definitions of $\beta$, $\gamma$, $\delta$)

$$x_1' = x_1 + \delta \cdot [\, 2x_1 + \beta \cdot (x_1+x_2) + \gamma \cdot (x_2^2 - x_1^2) - 4\,],$$
$$x_2' = x_2 + \delta \cdot [\, 2x_2 + \beta \cdot (x_1+x_2) + \gamma \cdot (x_1^2 - x_2^2) - 4\,].$$

Adding these two equations, we have (letting $X = x_1 + x_2$)

$$X' = X\{1 + 2\delta(1+\beta)\} - 8\delta.$$

Define

$$\delta' \equiv \delta(1+\beta); \quad k \equiv 2\delta/\delta' = 2/(1+\beta) = 2b_1/(b_1 + b_2); \quad Z \equiv X - 2k,$$

then we have

$$Z' = (1 + 2\delta')Z,$$

which implies (when $2\delta' < 1$)

$$X^{(m)} = (1+2\delta')^m (x_{10} + x_{20} - 2k) + 2k, \approx e^{2m\delta'} Z_0 + 2k,$$

where $Z_0 \equiv x_{10} + x_{20} - 2k$.

If instead of adding we subtract the two equations, we find ($Y \equiv x_2 - x_1$):

$$Y' - Y = 2\delta \cdot (1 - \gamma X) Y,$$

which implies

$$\ln Y' - \ln Y \approx 2\delta \cdot (1 - \gamma X).$$

or

$$\ln Y^{(m)} - \ln Y_0 \approx 2\delta \cdot \Sigma_{j=0,\ldots,m-1} (1 - \gamma X^{(j)}).$$

By substituting our previous solution for $X^{(j)}$, after some algebra we find

$$Y^{(m)} \approx Y_0 \cdot \exp[\, 2\delta m(1 - 2k\gamma) - \delta z_0 \gamma/\delta'(e^{2m\delta'} - 1) \,].$$

Defining

$$\phi \equiv \delta\,(1 - 2k\gamma) = 0.25 b_1/N\,(1 - 4c/(b_1 + b_2))$$

and

$$\theta \equiv \delta Z_0 \gamma/\delta' = (x_{10} + x_{20} - 2k) c\,/\,(b_1 + b_2),$$

we obtain finally:

$$X^{(m)} \approx e^{2\delta' m} \cdot Z_0 + 2k,$$
$$Y^{(m)} \approx (Y_0 \cdot e^{2\phi m}) \cdot \exp(-\theta[e^{2\delta' m} - 1]\,).$$

From the definitions of $X$ and $Y$ we have

$$(X+Y)/2 = x_2\,;\quad (X-Y)/2 = x_1,$$

and we may derive a vector form of the dynamical equation:

$$\mathbf{x}^{(m)} = e^{2\delta\,(1-2k\gamma)\,m} \cdot \exp(-\theta[e^{(1+\beta)\delta m} - 1]\,)\, q_1'[1\,;-1] + e^{2(1+\beta)\delta m}\, q_2'[1\,;1] + k\,[1\,;1],$$

where the constants $q_1'$ and $q_2'$ are solved from initial conditions. It is instructive to compare this result with the piecewise-linear model:

$$\mathbf{x}^{(m)} \approx e^{2\delta m} q_1[1;-1] + e^{2(1+\beta)\delta m}\, q_2[1 + s\,\gamma;\, 1 - s\,\gamma] + k[1 - s\,\gamma\,;\, 1 + s\,\gamma].$$

From comparing the two expressions, we find that they agree when $\gamma = 0$ as expected.

### 5.2.1.3 Effectiveness of "sports program" in sticky model

In Section 3 we described several sports programs that are intended to foster positive intergroup relations in divided communities. Let us consider the potential effectiveness of such programs under the sticky assumption, so that each group maintains unanimous affinities. In this case, we have shown that the value of $k \equiv 2b_1(b_1+ b_2)^{-1}$ is critical in determining the eventual affinities of both groups. Sports programs aim to create opportunities for positive interactions between groups: as such, they are targeted towards increasing $b_2$ for a subset of each group. Suppose that due to a single sporting event the mean value of $b_2$ for a proportion $p$ of each group's population is increased by a factor $r$. In this case, the average $b_2$ changes by: $b_2 \to b_2\{1+(r-1)p\}$. Then given that $(r-1)p <<1$, we have

$$k \equiv 2(1+ b_2/b_1)^{-1} \to 2 (1+ \{1 + (r-1)p\}b_2/b_1)^{-1} \approx k \{1 - (r-1)p[b_2/(b_1+ b_2)]\}.$$

It follows that the proportionate change in $k$ is $\{(r-1)p \cdot b_2/(b_1+ b_2)\}$, which is smaller than the mean percentage change in $b_2$ by a factor of $b_2/(b_1+ b_2)$. Thus if the within-group negative drift $b_1$ is large compared to the positive inter-group drift $b_2$, then the positive effect of the sports program is seriously eroded. This seems to imply that sports programs will be ineffective when strong negative within-group drifts exist. We will confirm this tentative conclusion with simulations in Section 6.

## 5.3 Case 2: polarized starting distributions

Rather than starting with unanimous groups as in Case 1, we may also consider the case where each group begins with extremist and moderate factions (with affinities 0 and 1, respectively). In the following, we consider the question of when such polarized-group distributions can remain stable.

It turns out that in this situation it is easier to work with the nonlinear model. We may follow the derivation of the dynamical equations for the sticky model, except instead of having two groups we have four: Groups 1a, 2a, 1b, 2b, which consist of extremists in Groups 1 and 2, and moderates in Groups 1 and 2, respectively. Groups 1a and 1b contain respectively fractions $1-f_1$ and $f_1$ of the agents in Group 1, and Groups 2a and 2b similarly contain fractions $1-f_2$ and $f_2$ of the agents in Group 2. Initially, we have characteristic affinities $x_{1a}, x_{2a} \approx 0$ and $x_{1b}, x_{2b} \approx 1$. With these four groups, using a procedure analogous to the derivation in the sticky model case, we obtain:

$$x_{1a}' \approx x_{1a} + (1/2N)(1-f_1)\{(1-f_1)(-2b_1)+f_1(-b_1 + c)\}+ (1/2N) \cdot f_1 \cdot \{(1/2)(1-f_1)(-b_1 + c)\}$$
$$+ (1/2N) \cdot f_2 \cdot \{(1/2)(1-f_1)(b_2 + c)\};$$

$$x_{1b}' \approx x_{1b} + (1/2N)(1-f_1) \cdot f_1 \cdot (-b_1 - c) + (1/2N) \cdot f_1 \cdot \{(1/2)(1-f_1)(-b_1 - c) + (1/2) \cdot f_1 \cdot (-2b_1)$$
$$+ (1/2)(1-f_2)(b_2 - c) + (1/2) \cdot f_2 \cdot b_2\} + (1/2N) \cdot f_2 \cdot \{(1/2) \cdot f_1 \cdot b_2\},$$

and similar equations with all '1' indices exchanged with '2'. The condition that $x_{1a}' \leq x_{1a}$ leads to:

$$f_2(b_2 + c) \geq -f_1(b_1 + 3c) + 4b_1,$$

while the condition that $x_{1b}' \geq x_{1b}$ leads to:

$$f_2(b_2+c) \geq -f_1(b_1 + 3c) + 3b_1 - b_2 + 4c.$$

Using the same derivation with the two groups exchanged, we obtain two additional equations with $f_1 \leftrightarrow f_2$. By comparing the two above inequalities, we obtain the following necessary conditions for stability:

$$c \leq (b_2+b_1)/4.$$

If $c$ does not satisfy this condition, then all polarized groups are unstable. From the second inequality we also obtain the necessary condition $b_2 \geq b_1$: if this condition is not satisfied, both groups will end up unanimously extremist.

## 5.4 Differential equation for evolution of probability densities

So far we have only analyzed cases where the distributions of affinities assume only a handful of values at any given time. But in a system with $N$ agents per group, it is actually possible for all $N$ agents to have different affinities at any given time. When $N$ is large, it is reasonable to expect that the distributions will become nearly continuous, and the behavior may be approximated by a differential equation. In the following we derive a differential equation for the evolution over time of the probability density, which can be solved exactly when the distributions' expected values are known functions of time. To do this, we perform a differential analysis of how individual interactions will affect the overall distribution.

Let $p_1(x_1)$ and $p_2(x_2)$ be the probability densities for the two groups, so that $\int p_1(x_1)dx_1 = \int p_2(x_2) dx_2 = 1$ and $0 \leq x_1, x_2 \leq 1$. From the above algorithm specification, we may find the differential transition probabilities into and out of the interval $[x_1, x_1+dx_1]$ during a single interaction between agents $A_1$ and $A_2$.

(I) The probability that the agent $A_1$ is in Group 1 and has affinity in $[x_1, x_1+dx_1]$ is $(1/2)p_1(x_1)dx_1$. This will cause $p_1(x_1)dx_1$ to decrease by $(1/N)$.

(II) The probability that agent $A_1$ and agent $A_2$ are in Group 1 and have affinities $u,v$ respectively such that $u-b_1+c(v-u) \in [x_1, x_1+dx_1]$ is

$$(1/2) \cdot \int dv\, p_1([x_1+ b_1-vc](1-c)^{-1}) \cdot (1-[x_1+ b_1-vc](1-c)^{-1}/2)\, p_1(v)\, (1-c)^{-1} dx_1.$$

This will cause $p_1(x_1)dx_1$ to increase by $(1/N)$.

(III) The probability that the agents $A_1$ and $A_2$ are both in Group 1, and $A_2$ has affinity in $[x_1, x_1+dx_1]$ is

$$(1/2) \cdot \int dv\, p_1(v)(1-v/2)\, p_1(x)\, dx_1.$$

This will cause $p_1(x_1)dx_1$ to decrease by $(1/N)$.

(IV) The probability that agent $A_1$ and agent $A_2$ are in Group 1 and have affinities $u, v$ respectively such that $v-b_1+c(u-v) \in [x_1, x_1+dx_1]$ is

$$(1/2) \cdot \int du\, p_1(u)(1-u/2)\, p_1([x_1+ b_1-uc](1-c)^{-1})\, (1-c)^{-1}\, dx_1.$$

This will cause $p_1(x_1)dx_1$ to increase by $(1/N)$.

(V) The probability that agent $A_1$ is in Group 1 and agent $A_2$ is in Group 2 with affinities $u, v$ respectively such that $u+b_2+c(v-u) \in [x_1, x_1+dx_1]$ is

$$(1/2) \cdot \int dv\, p_1([x_1- b_2-vc](1-c)^{-1})( [x_1- b_2-vc](1-c)^{-1}/2 )\, p_2(v)\, (1-c)^{-1}\, dx_1.$$

This will cause $p_1(x_1)dx_1$ to increase by $(1/N)$.

(VI) The probability that agent $A_1$ is in Group 2 and agent $A_2$ is in Group 1 with affinities $u, v$ respectively such that $v+b_2+c(u-v) \in [x_1, x_1+dx_1]$ is

$$(1/2) \cdot \int du\, p_2(u)\, (u/2)\, p_1( [x_1- b_2-uc](1-c)^{-1})\, (1-c)^{-1}\, dx_1.$$

This will cause $p_1(x_1)dx_1$ to increase by $(1/N)$.

(VII) The probability that agent $A_1$ is in Group 2 and agent $A_2$ is in Group 1 with affinities $u, v$ respectively such that $v \in [x_1, x_1+dx_1]$ is

$$(1/2) \cdot \int du\, p_2(u)\, (u/2)\, p_1(x_1)\, dx_1.$$

This will cause $p_1(x_1)dx_1$ to decrease by $(1/N)$.

We also suppose that $b_1, b_2, c \ll 1$ so we may use derivative approximations, and we may neglect quadratic terms in these variables. Among other simplifications, this enables us to replace $(1-c)^{-1}$ with $1+c$, So the contributions to $\Delta p_1(x_1)$ from the different terms are given by $(2N)^{-1}$ times the following factors:

(I) $-p_1(x_1)$

(II) $\int p_1(v) \cdot \{ p_1( (1+c)x_1+ b_1- vc)\, (1 - [(1+c)x_1+ b_1- vc]/2)(1+c) \} dv$

(III) $-\int p_1(v)(1-v/2)\, p_1(x)\, dv$

(IV) $\int p_1(v) \cdot \{ (1- v/2)\, p_1( (1+c)x_1+ b_1- vc )\, (1+c) \} dv$

(V) $\int p_2(v) \cdot \{ p_1( (1+c)x_1- b_2- vc )\, ( [(1+c)x_1- b_2- vc] /2)\, (1+c) \}\, dv$

(VI) $\int p_2(u) \cdot \{ (u/2)\, p_1( (1+c)x_1- b_2- uc)\, (1+c) \} du$

(VII) $-\int p_2(u)\, (u/2)\, p_1(x_1)\, du$

By integrating all of these terms over $x_1$ (making use of change of variable), we may verify that the overall change in $\int p_1(x_1)dx_1$ is 0, so that following the interaction the distribution still has total mass 1.

Using Taylor expansion, we have (for $c, k \ll 1$)

$$f(x+cx+k)(1+c) \approx f(x)(1+c) + f'(x)(cx+k)(1+c) \approx f(x) + c \cdot f(x) + (cx+k)f'(x)$$

or

$$f(x+cx+k)(1+c) \approx f(x) + [(cx+k)f(x)]',$$

where the prime (′) denotes derivative with respect to $x$. If we make this substitution in (I)–(VII), then all non-derivative terms and cancel and we end up with the equation:

$$\partial p_1/\partial t - p_1 K_1 = \partial p_1/\partial x_1 (K_1 x_1 + K_2),$$

with:

$$K_1 = 4c - b_1 - b_2; \quad K_2 = 4(b_1 - cE_1) - b_1 E_1 - b_2 E_2 + c(V_1 - V_2),$$

where $E_i$, and $V_j$ ($j = 1, 2$) are first and second moments respectively of the two groups:

$$E_1 = \int p_1(v) \cdot v \, dv; \quad E_2 = \int p_2(v) \cdot v \, dv; \quad V_1 = \int p_1(v) \cdot v^2 dv; V_2 = \int p_2(v) \cdot v^2 dv.$$

Notice that $E_j$ and $V_j$ depend on $t$ for $j=1,2$. However, if we assume that $E_j$ and $V_j$ are approximately constant, then the equation may be rewritten as

$$\partial/\partial t \, (e^{-K_1 t} p_1) = \partial/\partial y \, (e^{-K_1 t} p_1),$$

where

$$y = K_1^{-1} \cdot \log(K_1 x_1 + K_2) + C.$$

The solution is then ($q$ is an arbitrary once-differentiable function)

$$p_1 = e^{K_1 t} q(t+y) = e^{K_1 t} q(t + K_1^{-1} \cdot \log(K_1 x_1 + K_2) + C).$$

Defining $p_{10}(w) = q(\log(K_1 w)/K_1)$, we find:

$$p_1(x_1, t) = e^{K_1 t} p_{10} \cdot (e^{K_1 t}(x_1 + K_2/K_1)),$$

and a similar equation holds for $p_2(x_1, t)$ (with $E_1 \leftrightarrow E_2$ and $V_1 \leftrightarrow V_2$). Note that $p_{10}$ is a probability density, and is in fact a shifted version of $p_1(x_1, 0)$.

According to this solution, as the distribution of affinities evolves it maintains its basic initial shape, but becomes either stretched out or compressed over time, depending on whether $K_1 > 0$ or $K_1 < 0$, respectively. It also undergoes translation which varies according to the changes in first and second moments.

The above solution does not take into account the boundary conditions that prevent probability mass from escaping the interval [0,1], so it cannot accurately model systems in which probability mass moves to the boundaries. It does establish however that that no differentiable equilibrium solutions are possible when $K_1 \neq 0$. It also indicates the possibility of a point-mass solution at $x_1 = -K_2/K_1$, which implies (from the $p_1$ and $p_2$ equations, respectively)

$$(-4c + b_1 + b_2)x_1 = b_1(4-x_1) - b_2 x_2 + c(x_1^2 - x_2^2 - 4x_1),$$
$$(-4c + b_1 + b_2)x_2 = b_1(4-x_2) - b_2 x_1 + c(x_2^2 - x_1^2 - 4x_2).$$

After algebraic manipulation we obtain:

$$x_1 + x_2 = 4b_1/(b_1 + b_2) \quad \text{and} \quad 2(b_1 + b_2)(x_2 - x_1) = 8b_1 c(x_2 - x_1)/(b_1 + b_2),$$

so that either $x_2 - x_1 = 0$ or $c = (b_1 + b_2)^2/(4b_1)$. We will investigate the existence of such fixed points in the next section.

# 6  Simulations

In the previous section we derived mathematical equations to describe system behavior, under various particular conditions and relying on simplifying assumptions. In this section, we report on computer simulations which were conducted to ascertain the accuracy of these equations, and to further explore the general behavior of the model.

## 6.1 Unanimous-group starting conditions (Case 1)

In order to verify the accuracy of the "sticky assumption", we tracked the behavior of simulations where all individuals in Group $j$ starts with equal affinity $x_{j0}$ ($j$=1,2): this condition corresponds to Case 1, as described in Section 5.2. The entire region $(x_{10}, x_{20}) \in [0,1]^2$ was sampled, and the system parameters used are summarized in Table 1.

Table 1 Parameters in simulations for Case 1 (unanimous starting affinities for each group)

| Symbol | Significance | Value |
|---|---|---|
| $b_1$ | Negative drift from within-group interactions | 0.002 |
| $b_2$ | Positive drift from intergroup interactions | $3b_1$ |
| $c$ | Cohesion parameter | $0.25b_1$ (small $c$ case) |
| | | $4b_1$ (large $c$ case) |
| $N$ | Number of agents in each group | 250 |
| – | Number of iterations | $500N$ |
| $\sigma$ | Noise parameter in individual interactions | $2b_1$ (piecewise-linear model) |
| | | $b_1$ (nonlinear model) |
| $h(x)$ | Influence function | $\text{sign}(x)$ (piecewise-linear model) |
| | | $x$ (nonlinear model) |

### 6.1.1 Simulation for piecewise-linear model, $h(x) = \text{sign}(x)$

Simulation results for the case $h(x) = \text{sign}(x)$ (which gives rise to a piecewise-linear model, as described in Section 5) are shown in Figure 1. The two lines $x_1 = x_2$ and $x_1+x_2=1$ divide the square $[0,1]^2$ into quadrants, and the system exhibits symmetry about the $x_1=x_2$ line. (For general values $b_1$ and $b_2$, the dividing lines are $x_1+x_2=4b_1/(b_1+b_2)$ and $x_1=x_2$, and symmetry about $x_1=x_2$ is preserved.) Red asterisks indicate the different starting points used for different simulation scenarios: the $x_1$ (resp. $x_2$) coordinate corresponds to the starting affinity value for all members of Group 1 (resp. Group 2). Black lines show the subsequent trajectories predicted by the "sticky model" equations derived in Section 5.2.1.1. Blue and green dotted lines show respectively the trajectories followed by the ($10^{th}$, $10^{th}$) and ($90^{th}$, $90^{th}$) percentiles of affinities for (Group 1, Group 2) in the agent-based simulations. The red dotted line similarly shows the trajectory followed by the median values. Figure 1(*left*) corresponds to a small cohesion parameter ($c=0.25b_1$), while Figure 1(*right*) uses a large cohesion ($c=4b_1$), where $b_1$ is the within-group interaction drift.

In both the small-$c$ and the large-$c$ case, the sticky model trajectories derived in Section 5.2.1.1 predict quite closely the trajectories of the agent-based model medians. In the large $c$ case the $10^{th}$ and $90^{th}$ percentile trajectories remain close to the median trajectories, showing that the spreading of the distribution is limited which in turn implies that the sticky assumption is not unrealistic. In the small $c$ case the tracking is somewhat looser, especially in cases where the two groups' initial affinities are substantially different. The small $c$ figure shows one case with starting affinities $(x_1, x_2)$ on the $x_1+x_2=1$ line, in which the $10^{th}$ and $90^{th}$ percentiles tend towards (0,0) and (1,1), respectively. In this case, both populations are dividing into extremist and moderate factions. This situation is further explored in subsequent simulations.

In most cases the system tends either towards (0,0) or (1,1) (corresponding to all extremists and all moderates, respectively). When $c$ is small, in scenarios where the two populations beginning in the extreme upper left or lower right corners, one group may end up all nearly extremist (affinity≈0), while the other group ends up all nearly moderate (affinity≈1). When $c$ is large, in all cases the system move fairly directly towards the $x_1=x_2$ line, and the affinities in each group spread only slightly. Once the $x_1=x_2$ line is met, the system moves either towards (0,0) or (1,1): if the agent-based system meets the border of the square, it continues along the border until it reaches one of these corners. Since the size of the cohesion parameter $c$ reflects the degree to which interacting individuals influence each other, it stands to reason that when $c$ is large, the two groups should end up with the same overall affinity.

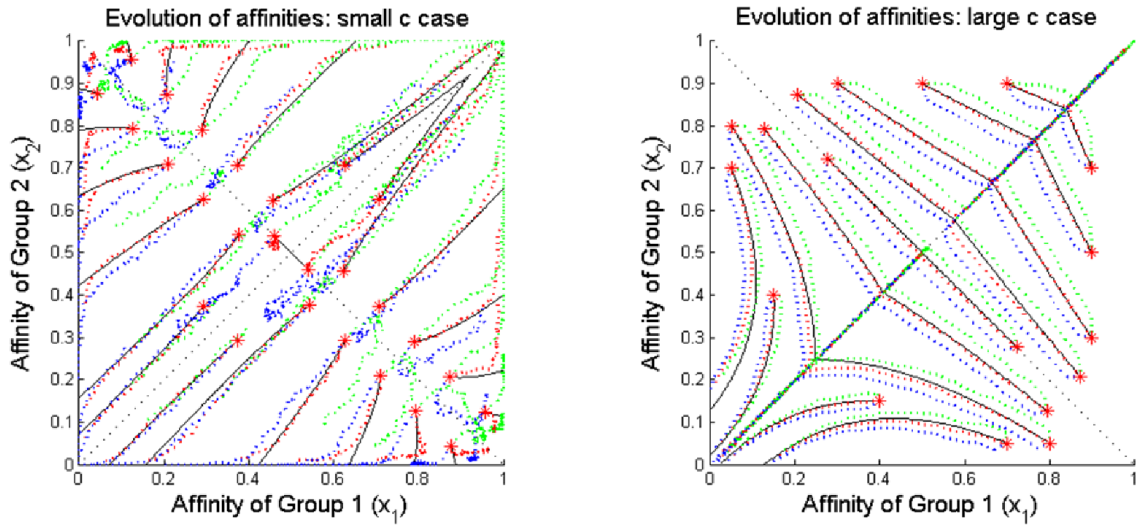

**Figure 1** Comparison of piecewise-linear "sticky model" predictions (with influence function $h(x)=\text{sign}(x)$) with agent-based simulations in which each group initially has unanimous affinities. Red asterisks show starting conditions; black trajectories show sticky-model theoretical predictions; and blue, red, and green dotted lines show the trajectories of $10^{th}$ percentiles, medians, and $90^{th}$ percentiles as explained in the text. Figure (a) uses a value of the cohesion parameter $c$ that is 1/4 times the intergroup drift $b_1$, while the figure (b) uses $c=4b_1$.

### 6.1.2 Simulation of nonlinear model, $h(x) = x$

System trajectories for a linear influence function $h(x)=x$ (which gives rise to a nonlinear model), are shown in Figure 2. Most of the characteristics of the piecewise-linear model described above also apply to the nonlinear model. Once again, the agent-based model medians follow quite closely the theoretical sticky-model trajectories (derived in Section 5.2.1.2). The spreading of affinity distributions (evidenced by the divergence of blue and green dotted trajectories) is somewhat greater than in the piecewise-linear case (the noise factor $\sigma$ was reduced by half to reduce the spreading and make trajectories clearer.)

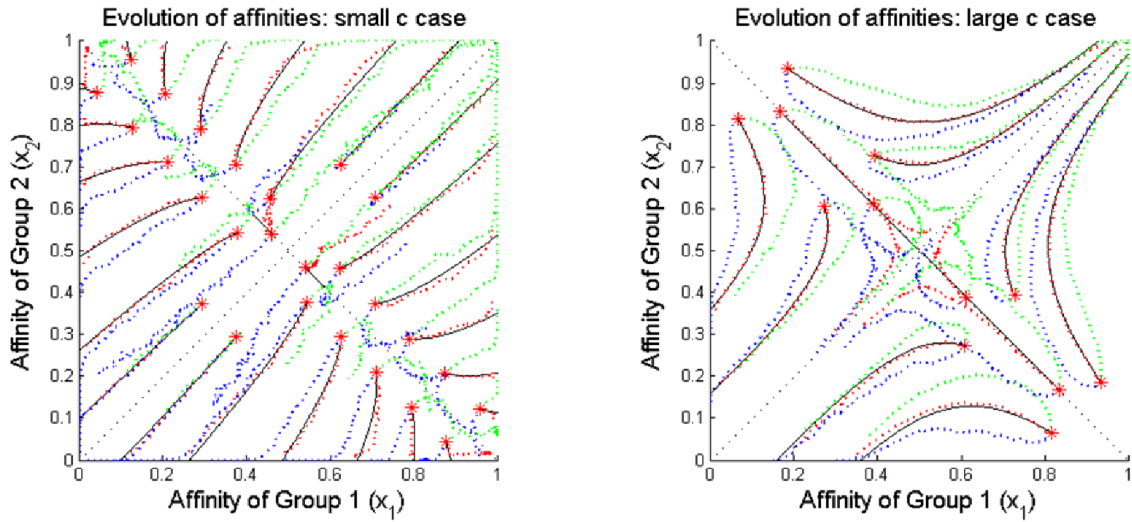

**Figure 2** Comparison of nonlinear "sticky model" predictions (with influence function $h(x) = x$) with agent-based simulations in which each group initially has a single characteristic affinity. Trajectories have the same interpretation as in Figure 1.

We conclude from both piecewise-linear and nonlinear model simulations that the "sticky model" is quite accurate in cases where all individuals in each group begins with the same opinion of the other group. It is especially accurate if the cohesion parameter $c$ is relatively large. Apparently these results are not particularly sensitive to the form of the influence function $h$. Except for extreme cases, the eventual fate of the system depends entirely on the position of initial $(x_1, x_2)$ compared with the line $x_1 + x_2 = 4b_1(b_1 + b_2)^{-1}$. If the initial position lies below (resp. above) the line, then all individuals become extremists (resp. moderates). Only if the initial position lies exactly on this line is it possible for groups to split into extremist and moderate factions. The tendency to split is enhanced when the cohesion $c$ is smaller.

These results seem to indicate that early intervention is absolutely critical in the case of divided communities with the potential of developing antagonistic relations. Intervention should attempt to reduce negative drifts produced by within-group interactions, as well as encourage positive interactions between groups. There appears to be a minimum level of effective intervention, below which the situation will progressively deteriorate and above which the situation will eventually achieve universal moderation.

## *6.2 Sports program simulations*

We may simulate sports programs such as described in Section 3 by modifying the model to include occasional "sporting events" which involve a select group of agents, half from each group. Sporting events occur randomly at a given frequency, and each sporting effect improves the affinity of all participants by a determined amount $b_{2+}$. Parameters for the simulations are shown in Table 2. Figure 3 shows the equilibrium proportion of moderates (color scale) as a function of the initial affinity (all individuals in both groups begin with the same affinity) and fraction of sports program participants from each group.

**Table 2 Parameters for sports program simulations**

| Symbol | Significance | Value |
|---|---|---|
| $b_1$ | Negative drift for within-group interactions | 0.003 |
| $b_2$ | Positive drift for between-group interactions | 0.007 |
| $c$ | Cohesion parameter | $b_1/3$ (small $c$); $3b_1$ (large $c$) |
| $b_{2+}$ | Increase in affinity for sports participants | 0.003 |
| – | Frequency of sports programs | 0.01 |
| – | Fraction of each group that participate | 0.05–0.95 |
| – | Starting affinity (unanimous) | 0.05–0.95 |
| $N$ | Number of agents in each group | 250 |
| – | Number of iterations | $2000N$ |
| $\sigma$ | Noise parameter in individual interactions | 0 |
| $h(x)$ | Influence function | $x$ (nonlinear model) |

Figure 3 *(left)* shows the situation when cohesion between individuals in the society is small ($c=b_1/3$). In this case, when the initial unanimous affinity falls below a certain threshold the effectiveness of sports programs is limited to program participants only. For example, when the participation rate is 0.1 then the threshold is roughly 0.65: this means that if the initial unanimous affinity is less than 0.65, then only program participants become moderates and the remainder of the population is relatively unaffected (in the figures, "moderate" is defined as possessing an affinity above 0.9). On the other hand, if the initial unanimous affinity is above the threshold, the influence of the program is propagated throughout both groups, and virtually 100% of both groups attain moderate affinities. The threshold decreases gradually as the proportion of participants increases.

Figure 3 *(right)* shows the situation when cohesion between individuals in the society is large ($c=3b_1$). In this case, the effectiveness of sports programs is greatly enhanced compared to the small-$c$ case. For low rates of participation, thresholds still exist below which only participants become moderate in equilibrium: but these thresholds are lowered compared to the small-$c$ case, and only exist when the rate of participation is under about 20 percent. In all other cases, moderation pervades the entire population.

These results point to the effectiveness of sports programs (or similar programs designed to produce positive interactions between subgroups) when the relational situation has not degenerated too seriously. The effectiveness of such programs is greatly reduced when cohesion within the population is reduced.

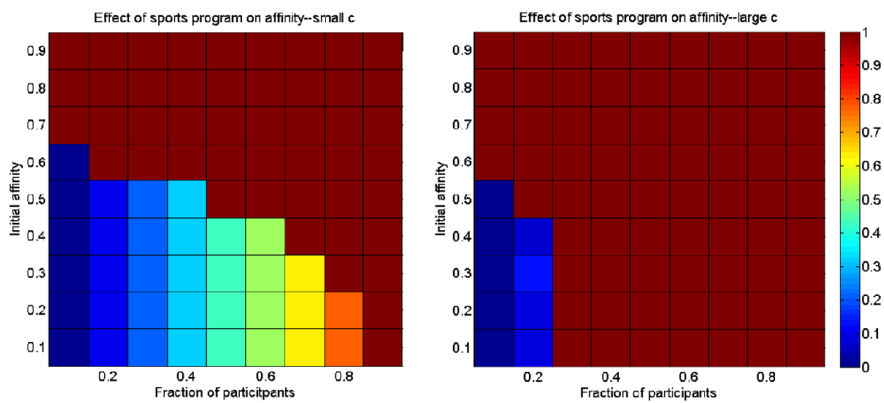

**Figure 3 Equilibrium proportions of moderates (affinity > 0.9) for communities with simulated sports programs for $c = b_1/3$ (left) and $c = 3b_1$ (right). The equilibrium affinity for each (fraction of participants, initial affinity) combination is indicated by the color scale running from 0 (deep blue) to 1 (dark red).**

## 6.3  Polarized starting conditions (Case 2)

Simulations were performed to verify the theoretical predictions made in Section 5 for the case where both groups are initially divided between extremist (affinity=0) and moderate (affinity=1) fractions (which we designated as Case 2). Parameters used in the simulation are listed in Table 3.

**Table 3 Parameters for polarized-group simulations (Case 2)**

| Symbol | Significance | Value |
|---|---|---|
| $b_1$ | Negative drift for within-group interactions | 0.003 |
| $b_2$ | Positive drift for between-group interactions | 0.007 |
| $c$ | Cohesion parameter | 0.001 |
| $f_1$ | Starting fraction of moderates (Group 1) | $0 < f_1 < 1$ |
| $f_2$ | Starting fraction of moderates (Group 2) | $0 < f_2 < 1$ |
| $N$ | Number of agents in each group | 250 |
| – | Number of iterations | $4000N$ |
| $\sigma$ | Noise parameter in individual interactions | 0 |
| $h(x)$ | Influence function | $x$ (nonlinear model) |

Figure 4 shows the theoretical region of stable $(f_1, f_2)$ configurations for the given set of parameters, where $f_j$ represents the starting proportion of moderates in group $j$, $j=1,2$.

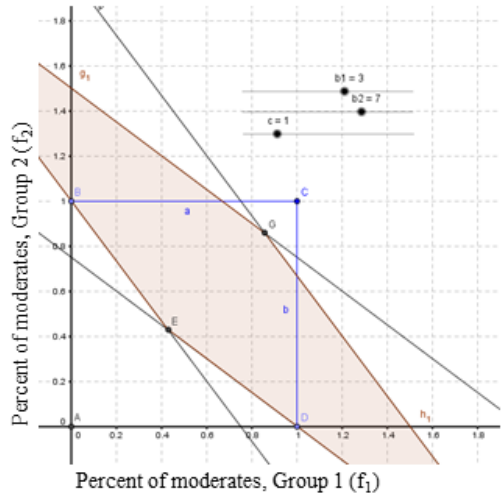

**Figure 4 Theoretical region of polarized stability, as a function of initial moderate percentages. (The figure was generated using the open-source software Geogebra.)**

Simulation results are depicted in Figure 5 (*left*) and (*right*), which show the equilibrium proportion of moderates (affinity > 0.9) in Groups 1 and 2 respectively, as a function of different $(f_1, f_2)$ starting configurations. The dark blue regions correspond to initial $(f_1, f_2)$ configurations that eventually lead to all extremists; while the dark red regions correspond to initial configurations that eventually lead to all moderates. The in-between rainbow-colored region, which resembles the theoretical region in Figure 4, give stable $(f_1, f_2)$ pairs. The vertical color stripes in the figure at left show that the equilibrium moderate fraction for Group 1 is nearly independent of the starting $f_2$ value; and the horizontal stripes in the figure at left indicate that the same statement is true with 1↔2. This confirms that the region of $(f_1, f_2)$ stability for the agent-based model conforms closely to the theoretical prediction. Note the region is actually somewhat smaller than theory would suggest, reflecting the fact that the agent-based dynamics produces fluctuations that undermine the stability of the system.

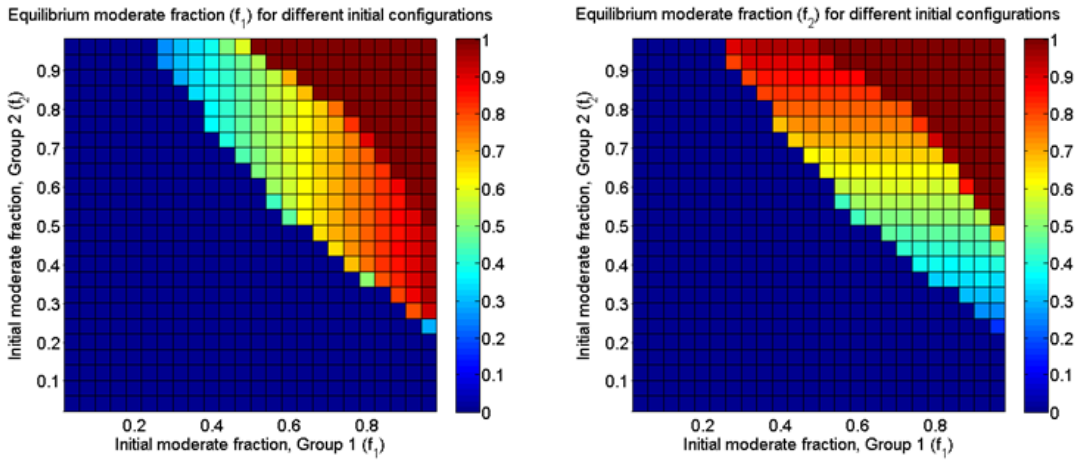

**Figure 5 Equilibrium moderate fractions for Group 1 (*left*) and Group 2 (*right*) as a function of starting (Group 1, Group 2) moderate fractions in Case 2. The equilibrium moderate fractions are indicated by the color scales at right.**

## 6.4 Uniform starting distributions (Case 3)

We also simulated the case where both groups start from uniform distributions, using the parameters in Table 4. The ratios $b_2/b_1$ and $c/b_1$ may be considered as the independent parameters which determine the eventual behavior of the system. (Changing the value of $b_1$ while holding the ratios $b_2/b_1$ and $c/b_1$ will only change the

effective time scale over which changes occur, but will not affect the eventual equilibrium values.)

**Table 4 Parameters for simulations starting from uniform distribution**

| Symbol | Significance | Value |
|---|---|---|
| $b_1$ | Negative drift for within-group interactions | 0.01 |
| $b_2$ | Positive drift for between-group interactions | $0.18b_1 \leq b_2 \leq 5.6b_1$ |
| $c$ | Cohesion parameter | $0.18b_1 \leq c \leq 5.6b_1$ |
| $N$ | Number of agents in each group | 250 |
| – | Number of iterations | $2000N, 200N$ |
| $\sigma$ | Noise parameter in individual interactions | 0.01 |
| $h(x)$ | Influence function | $x$ (nonlinear model) |

Figure 6 shows the equilibrium 10[th] percentile, median, and 90[th] percentiles (*center, left,* and *right* respectively) using a color scale, as a function of the parameters $c/b_1$ and $b_2/b_1$. The three graphs are virtually identical, which means that all individuals eventually arrive at the same affinity. The graphs also show the final value is independent of $c$. Furthermore, there is a cutoff value of $b_2/b_1$ below which unanimous extremism is attained, and above with unanimous moderation is the result. The cutoff corresponds to $\log_{10}(b_2/b_1) \approx 0.47$, or $b_2 \approx 3b_1$: as we have seen in Section 5.2, this corresponds to the situation shown in Figure 1 and Figure 2, where the line $x_1+x_2=1$ is a critical line which divides starting conditions that go to unanimous extremism from those that go to unanimous moderation. In particular, if $x_1=x_2=x$ is the unanimous starting affinity, then $x=1/2$ is the critical value below which the system ends up in unanimous extremism, and above which all become moderates. It follows that if $b_2 < 3b_1$, then a system that starts unanimously with $x_1=x_2=1/2$ will end up in extremism; while if $b_2 > 3b_1$, then the same starting conditions will lead to unanimous moderation. In other words, the eventual fate of the system with uniform starting distribution of affinities is the same as a system where all individuals begin with affinity 1/2, which is both median and mean of the uniform distribution.

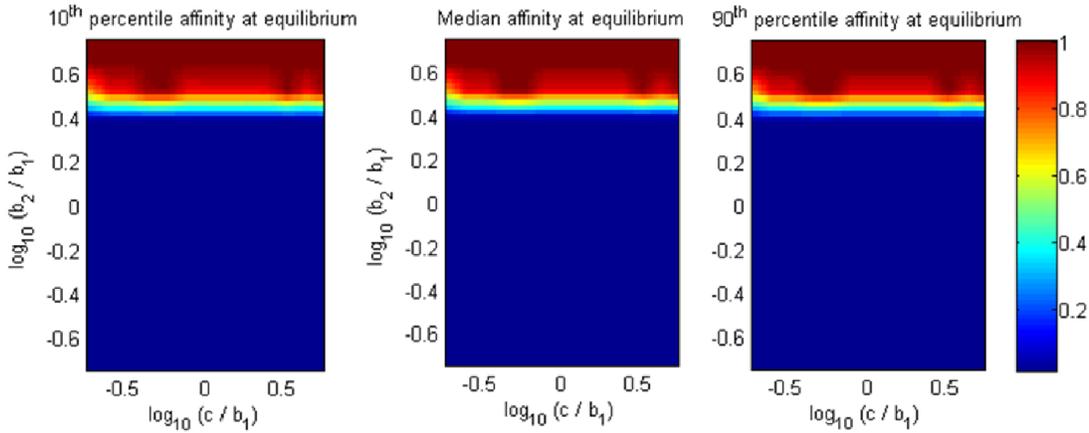

**Figure 6 Equilibrium 10[th] percentile, median, and 90[th] percentile affinities for system starting from uniform distribution, as a function of system parameter ratios $c/b_1$ and $b_2/b_1$. System parameters are given in Table 4. Affinity values are according to the color scale at right.**

Figure 7 gives information about the speed of convergence of the system with uniform starting distribution. System parameters are the same as in Figure 6, but the simulation is run for fewer iterations. The results show that 10[th] percentile values converge less rapidly when $\log_{10}(b_2/b_1)\approx 0.47$ and $\log_{10}c/b_1<0$. In Section 5.4 we saw that the sign of $K_1 = 4c - b_1 - b_2$ determined the convergence/divergence behavior of the theoretical continuous solution. In the current simulation, $K_1=0$ corresponds to $c=0.875b_1$, or $\log_{10}c/b_1=-0.06$. It follows that a change in the sign of $K_1$ does indeed correspond to a discontinuous change in the convergence behavior, as might be expected from the theoretical discussion. We will explore this connection further in the next section.

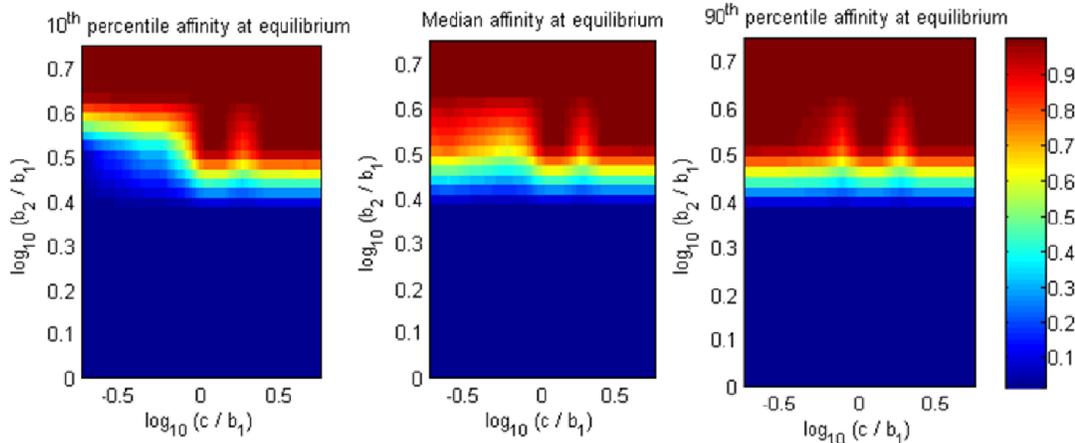

**Figure 7 Equilibrium median affinities for systems starting from uniform distribution. All system parameters are as in Figure 1, but the number of iterations is reduced to 200$N$. Median and 90$^{th}$ percentiles have converged in this time frame, but 10$^{th}$ percentiles are slower to converge for systems with log$_{10}$ $c/b_1$ < −0.06 and log$_{10}(b_2/b_1)$ ≈ 0.47.**

## 6.5   Spreading dynamics of distribution as a function of cohesion.

In Section 5.4 we inferred the possible existence of fixed-point solutions for the affinity distributions in which all affinities are identical, for certain parameter values. In this section we numerically investigate the behavior when the system is started at one of these potential fixed points. Table 5 shows parameters for these simulations: the values of $x_1$ $x_2$ were fixed at $2b_1 / (b_1 + b_2)$ in accordance with the potential fixed point values computed in Section 5.4.

**Table 5 Parameters for simulations starting from uniform distribution**

| Symbol | Significance | Value |
|---|---|---|
| $b_1$ | Negative drift for within-group interactions | 0.01 |
| $b_2$ | Positive drift for between-group interactions | $2.5b_1$ |
| $c$ | Cohesion parameter | $b_1/3$ and $b_1$ |
| $N$ | Number of agents in each group | 250 |
| – | Number of iterations | 1000$N$ |
| $\sigma$ | Noise parameter in individual interactions | 0.0 |
| $h(x)$ | Influence function | $x$ (nonlinear model) |
| $x_1, x_2$ | Unanimous starting affinity | $2b_1 / (b_1 + b_2)$ |

Figure 8 shows results of two typical simulations when the cohesion is small ($c=b_1/3$). For each iteration number (*x*-axis scale), the affinities (*y*-axis scale) of individuals in Groups 1 and 2 are plotted as green and red dots respectively. Simulations invariably end up either all extremist or all moderate, and the result varies from simulation to simulation. The distributions bifurcate temporarily into extremist and moderate factions, with relatively few individuals having intermediate affinities. This bifurcated state lasts between 200-400 iterations, after which all affinities either migrate upwards or downwards.

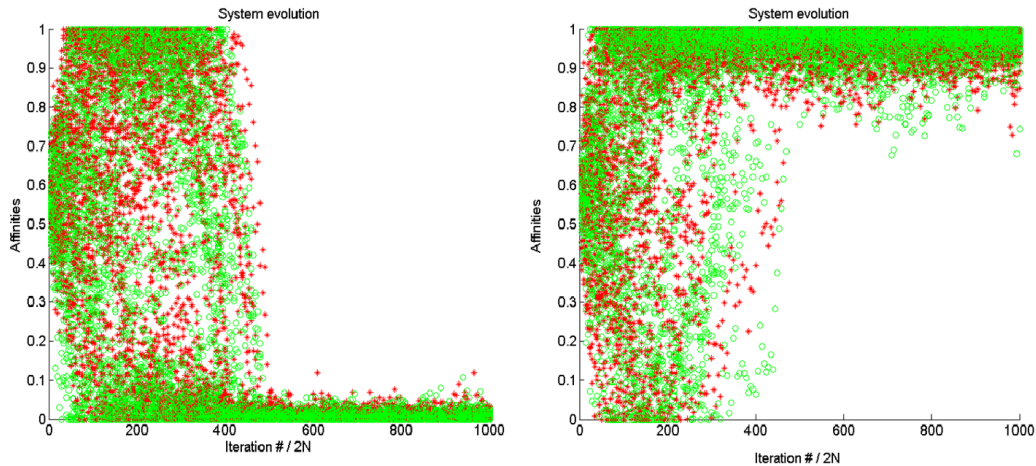

**Figure 8 Example distribution evolutions for system starting from theoretical fixed point, small-cohesion case ($c = b_1/3$). Group 1 and Group 2 individuals are represented by green and red dots, respectively.**

Figure 9 shows example system histories for a larger values of $c$ ($c=b_1$). Once again, affinities eventually become unanimously extremist or unanimously moderate, depending on the particular simulation. In this case however, there is no bifurcation. This result is consistent with the theoretical result derived in Section 6.5 that distribution spreading should be expected when $4c - b_1 - b_2 < 0$, while distributions should tend to stick together when $4c - b_1 - b_2 > 0$.

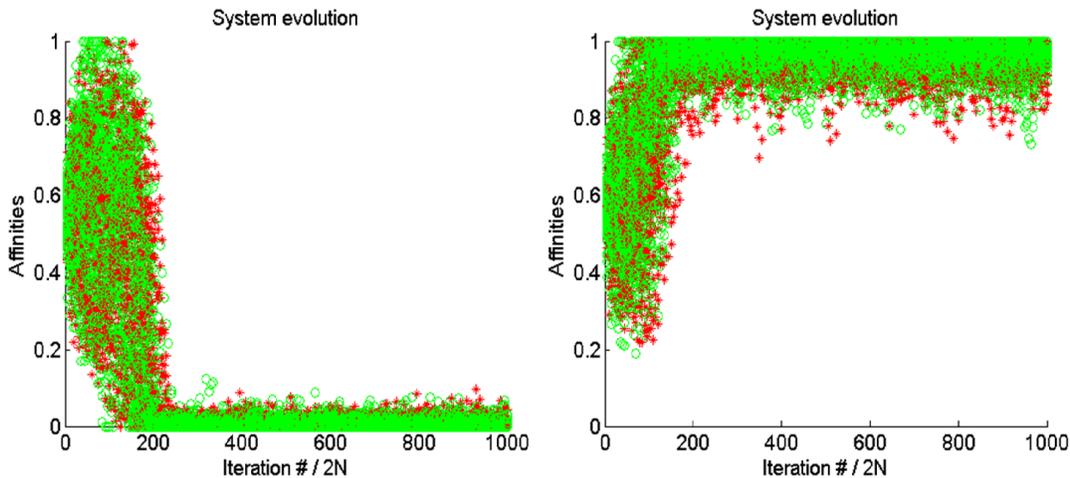

**Figure 9 Equilibrium distribution starting from theoretical fixed point, medium-cohesion case ($c=b_1$)**

These conclusions are confirmed in Figure 10, which represents the dispersive behavior starting from the theoretical fixed point as a function of $c$. Each vertical blue segment corresponds to a single simulation: lower and upper endpoints of the segment correspond to the final 10th and 90th affinity percentiles, respectively, while the dot within the segment denotes the final median affinity. Each segment is located above the corresponding simulation's value of $\log_{10}(c/b_1)$, shown on the $x$ axis. System parameters are as in Table 5, except that $c/b_1$ is allowed to vary as shown on the $x$-axis scale. Figure 10 (*left*) shows simulation results after $250N$ iterations, while the Figure 10 (*right*) shows results after $1000N$ iterations. The green line in each figure shows a local Gaussian-smoothed average of the median affinities for simulations with nearby values of $c/b_1$.

The figures shows that lower values of $c$ are associated with increased spreading, slower convergence, and a greater tendency to end up in extremism. There is a fairly rapid changeover in behavior in the interval $-0.2 < \log_{10}(c/b_1) < 0$, which is close to theoretical changeover point $\log_{10}(c/b_1) \approx -0.06$ corresponding to the condition $4c - b_1 - b_2 > 0$. When $c$ becomes larger, the percentages of simulations that end up extremist/moderate tend towards 50/50 (as shown by the green line).

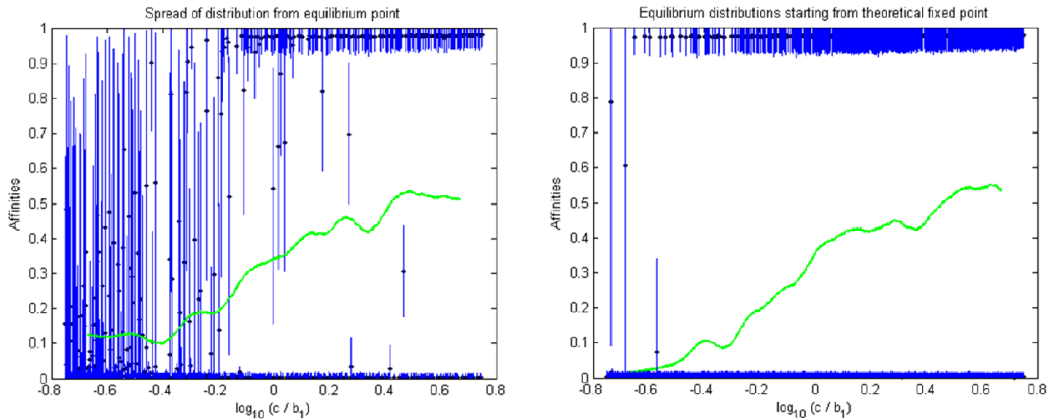

**Figure 10** Final 10[th] percentiles, medians, and 90[th] percentiles for affinities for multiple simulations starting from theoretical fixed point, for a range of values of $c/b_1$.

# 7   Discussion and conclusions

The theoretical and simulation results presented above have potentially important implications for policy decisions for dealing with divided communities with the potential for intergroup violence.

Both theory and simulations point out the unstable nature of the system. Social pressures produce polarization, and individuals either tend increasingly towards moderation or extremism. The middle ground eventually vanishes.

To achieve a moderate society, a much stronger positive tendency due to intergroup interactions is required to overcome the negative tendency associated with within-group interactions.  In practice, positive drift may result from mutual educational, economic, cultural and/or social advantages derived from intergroup interactions.  Any positive gains in affinity can be erased by negative between-group interactions (such as exploitation or hooliganism) that reduce or negate the between-group drift parameter $b_2$.

The model indicates that rapid intervention is required to reverse negative trends, because such reversals rapidly become increasingly more difficult. There exists a threshold level of intervention, below which the intervention is ineffective and may be counterproductive, and above which the desired goal of moderation will eventually be achieved.

Simulations indicate the effectiveness of sports programs when relationships between groups have not seriously degenerated. But if antipathy is already strong, sports programs may actually worsen the situation by singling out participants and creating distance between them and others in their own community.  Effectiveness of sports programs are seriously undermined if social cohesion is weak.

It is possible for a divided population to have stable moderate and extremist factions in each group. However, some of these cases are marginally stable, and a very slight change in overall affinity can push the system over the edge to become all moderate or all extremist, depending on the original configuration of factions.

For continuous affinity distributions, even if the distribution is quite disperse (as in the case of a uniform distribution) the tendency is for the system to end up either with unanimous moderation or unanimous extremism. The major determinants of the system's eventual fate appears to be the sum of median affinities of the two groups, and the drifts associated with within-group and intergroup interactions.

The degree of cohesion present in the society has significant influence on the dynamics. There is not a sharp critical level of cohesion where the behavior suddenly changes, but there is a fairly narrow changeover interval: the location of this interval on the negative within-group interaction drift and the positive between-group interaction drift. Communities in which cohesion is weakened are vulnerable to polarization, and are more difficult to reach through community outreaches that target only a portion of the population. Our study suggests that increased extremism in recent years may be  related to reduced interpersonal cohesion within groups resulting from modernization.